\documentclass{article}
\usepackage{chicago}
\usepackage{latexsym}

\newcommand{\iec}{\mbox{i.\,e.\,}}

\newcommand{\egc}{\mbox{e.\,g.\,}}

\newcommand{\dbd}[2]{\ensuremath{\frac{\dr{#1}}{\dr{#2}}}}

\newcommand{\mc}[1]{\ensuremath{\mathcal{#1}}}
\newcommand{\dr}[1]{\ensuremath{\mathrm{d} #1\,}}

\newcommand{\be}{\begin{equation}}
\newcommand{\ee}{\end{equation}}

\begin{document}

\title{Gravity, entropy, and cosmology: in search of clarity}
\author{David Wallace}
\date{June 2009}
\maketitle

\begin{abstract}
I discuss the statistical mechanics of gravitating systems and in particular its cosmological implications, and argue that many conventional views on this subject in the foundations of statistical mechanics embody significant confusion; I attempt to provide a clearer and more accurate account. In particular, I observe that (i) the role of gravity \emph{in} entropy calculations must be distinguished from the entropy \emph{of} gravity, that (ii) although gravitational collapse is entropy-increasing, this is not usually because the collapsing matter itself increases in entropy, and that (iii) the Second Law of Thermodynamics does not owe its validity to the statistical mechanics of gravitational collapse.
\end{abstract}

\section{Introduction}

On pain of a crisis in physics, the entropy of the early Universe had better be lower than its present entropy. This is sometimes presented as some deep requirement at the core of statistical mechanics, but there are more straightforward reasons why it had better be true: namely, the Second Law of thermodynamics says that the entropy of a closed system never decreases (and has increased in a great many processes which have occurred in the past and continue to occur today); the early Universe is in the past; therefore if its entropy is higher than the present Universe's entropy, the Second Law has been violated.

And in fact, it is not merely necessary that the entropy of the early Universe was lower than at present: it must have been lower in such a way as to facilitate the existence of the multifarious out-of-equilibrium systems (from stars to ice cubes) that we observe around us. If total entropy was lower in the past than in the present only because some distant region of the Universe had enormously increased in entropy, this would not facilitate the existence of out-of-equilibrium processes in the terrestrial neighborhood. It must have been the case, rather, that \emph{typical} regions of the Universe had lower entropy in the past than they do today.\footnote{As a matter of logic, typicality is not required: it would suffice that \emph{some} regions had lower entropy, whereupon we could appeal to anthropic arguments to explain why ours is such a region. (\citeN[p.419]{earmanph} seems to suggest something like this.) But of course, as a matter of observed fact there is plenty of evidence --- notably, the fact that the stars shine --- to show that entropy-increasing processes are ubiquitous across the Universe, and not just a local anomaly.}

Supposedly, there is a problem --- or at any rate, a prima facie problem --- with all this. For we know rather a lot, through more and less direct observations, about the state of the early Universe, and it seems to have been a very hot, very uniform gas. 

Why is this a problem? Sometimes one has the impression that it is because very hot, very uniform gases are not ``low-entropy'' states. But this is silly: `low' is a relative term. Whatever temperature the Universe had at a given size (more precisely, at a given baryon density), it would have been higher-entropy if it had been hotter, and lower-entropy if it had been colder. 

A slightly better reason to worry is that the early Universe seems to have had a higher entropy than the present-day Universe, which after all is not a very hot, very uniform gas. But a moment's thought shows that this is at best not obvious. The early Universe was very hot, and also very dense; as it expanded and cooled, it became less hot, and less dense. Since entropy is a measure of phase-space volume, ceteris paribus cooling lowers entropy and expanding raises it, and it isn't immediately obvious which should dominate.

The genuinely cogent worry is not that the early Universe was in a high-entropy state, nor that it had higher entropy than the present Universe, but that it is apparently in thermal equilibrium, and hence in the maximum-entropy state available to it; this, it would seem, is incompatible with the fact that the present-day Universe is not in thermal equilibrium. Let us call this the \emph{Initial State Problem}.

There is a proposed approach to this problem which has become extremely popular in the literature on the foundations of statistical mechanics. Its claim: the assumption that uniformity equates to high entropy ignores the existence of gravitation. Given the attractive nature of gravity (it is claimed), a uniform state is actually much lower-entropy than a much more clumped state.  Various forms of argument are given for this; frequently, it is argued that when attractive long-range forces are present, matter has a higher entropy when highly concentrated than when diffuse, and black holes (with their well-known stratospheric entropies) are exhibited as a limiting case.

In this paper, I wish to argue that this proposed approach, at least as it seems to be understood by most writers on the subject, contains a good deal of confusion. I will discuss the nature of statistical mechanics for gravitating systems, and explain in what sense it differs from non-gravitating systems, and I will argue that black holes are less of a limiting case and more of a special case. I will then move on to the thermodynamics of the early Universe, and make the case that in fact the local validity of the Second Law, and the solution to the Initial State problem, have much less to do with gravitational issues than has been thought. Although most or all of my account is really just a discussion of known physics, there is --- I hope --- clarity to be obtained from the conceptual issues brought up in that discussion.

\section{Three principles to set aside}\label{principles}

For the sake of a clearer debate, I will first state three 
principles which seem to underlie much of the gravitation-based approach to the Initial State Problem. I come to bury these principles, not to praise them: I will argue that each is at least misleading, and in general simply false. I will quote various authors in defence of these principles, if only to establish that I am not criticising straw men. However, my objective here is not to criticise specific authors' overall views on the topics of gravitational entropy and cosmology, but simply to show that many commonly accepted starting points for discussion are less stable and safe than is generally recognised.

\begin{description}
\item[First conventional principle:] ``Taking account of gravity'' in a discussion of entropy means allowing for ``gravitational entropy'', the entropy of the gravitational degrees of freedom. In particular, the early Universe had very high entropy in its non-gravitational degrees of freedom but very low gravitational entropy.
\end{description}
For an example of this view in print, consider Joel Lebowitz, speaking approvingly of a discussion by Roger Penrose:
\begin{quote}
He takes for the initial macrostate of the universe the smooth
energy density state prevalent soon after the big bang: an equilibrium state
(at a very high temperature) except for the gravitational degrees of freedom
which were totally out of equilibrium, as evidenced by the fact that the
matter-energy density was spatially very uniform. \cite[p.19]{lebowitz07}
\end{quote}

\begin{description}
\item[Second conventional principle:] When we do take account of gravity, dispersed low-density systems have low entropy, and concentrated high-density systems have high entropy. The stupendous entropy of black holes is simply a limiting case of this process.
\end{description}
David Albert, for instance, writes
\begin{quote}
Consider (for example) how stars come into being. What happens (we think) is that an initially dispersed cloud of dust\footnote{By `dust', Albert actually means (or should mean) gas: the interstellar medium from which stars form consists largely of atomic and molecular hydrogen.} --- under the influence of its own gravitation --- \emph{clumps up}. And it is as clear as it can be that the clumping up is a thermodynamically \emph{irreversible} process. And so it must be the case (and here is something unfamiliar from our considerations of gasses --- here is the influence of \emph{gravity}) that the entropy of the clumped-up condition is \emph{higher} than the entropy of the dispersed one. It must be (that is) that the dust particles are overwhelmingly likely to \emph{pick up a great deal of momentum}, in all sorts of different directions, as they fall in toward one another --- it must be (that is) that the clumped-up condition is the condition that's nonetheless by far the more \emph{dispersed} one in the [phase] space. \cite[p.90]{alberttimechance}
\end{quote}
Sheldon Goldstein puts it more succinctly:
\begin{quote}
[T]he attractive nature of the gravitational interaction is such that gravitating matter tends to clump, clumped states having larger entropy. \ldots For an ordinary gas, increasing entropy tends to make the distribution more uniform. For a system of gravitating bodies the reverse is true. High entropy is achieved by gravitational clumping --- and the highest of all, by collapse to a black hole.
\cite{goldsteinboltzmann}
\end{quote}
\begin{description}
\item[Third conventional principle:] Without the entropy-increasing effect of gravitational clustering, there would be no 2nd Law of thermodynamics. In particular, systems in our local neighborhood which are out of equilibrium have got that way because the process of gravitational clustering has allowed their entropy to drop.
\end{description}
Defenders of this principle have included Roger Penrose:
\begin{quote}
We seem to have come to the conclusion that all the remarkable lowness of entropy that we find about us --- and which provides this most puzzling aspect of the second law --- must be attributed to the fact that vast amounts of entropy can be gained through the gravitational contraction of diffuse gas into stars. Where has all this diffuse gas come from? It is the fact that this gas starts off as \emph{diffuse} that provides us with an enormous store of low entropy. We are still living off this store of low entropy, and will continue to do so for a long while to come.
\ldots
We have found where the diffuse gas has come from. It came from the very fireball which was the big bang itself. The fact that this gas has been distributed remarkably uniformly throughout space is what has given us the second law --- in the detailed form that this law has come to us --- after the entropy-raising procedure of gravitational clumping has become available.\cite[pp.\,417,425]{penroseenm}
\end{quote}
And Huw Price:
\begin{quote}
[The uniform early Universe] is the \emph{only} anomaly necessary to account for the vast range of low-entropy systems we find in the universe. In effect, the smooth distribution of matter in the early universe provides a vast reservoir of low entropy, on which everything else depends. \cite[p.228]{priceinhitchcock}
\end{quote}
The Third Principle, in turn, implies a Conventional Solution to the Initial State Problem:
\begin{description}
\item[Conventional solution:] Although the early Universe was at local thermal equilibrium, it was not at global thermal equilibrium because it was highly uniform and the process of becoming non-uniform is entropy-increasing when gravity is taken into account.
\end{description}
Before explaining just why these principles, and the Conventional Solution, are at best partially correct --- and in general, simply misleading --- I should mention a fourth principle:
\begin{description}
\item[Fourth principle:] Once one allows for gravitational entropy, the uniform early Universe can be seen to have such low entropy --- and thus, to have such an unlikely microstate --- that it is an urgent question for physics to ask just why this is.
\end{description}
This fourth principle has been strongly advocated by Penrose~\citeyear{penroseenm,penroseroadtoreality}, espoused by (e.g.) Price~\citeyear{priceinhitchcock,pricetimesarrow}, and criticised by (e.g.) \citeN{callenderinhitchcock} and \citeN{earmanph}. In this paper, though, I will be silent on the matter: my concerns are different.

\section{The irrelevance of ``gravitational entropy''}

The first issue I wish to address --- and the one which bears on my ``first principle'' is that of gravitational entropy: namely, is it true that to allow for gravity in considering the thermodynamics of a system is to allow for the entropy of that system's gravitational degrees of freedom?

It will help here if we consider a simple model (the first of many). Suppose that we have a gas of charged particles --- a plasma, in other words --- and wish to understand its thermodynamics. We analyse it mathematically and are feeling pleased with our analysis; then a colleague informs us smugly that we have failed ``to allow for electromagnetism''. What could this mean?

Well, there are two aspects of this problem. Firstly, since the particles in the plasma are charged, they will interact with the electromagnetic field. At equilibrium, then, the system will consist both of plasma particles at temperature $T$ and of black-body electromagnetic radiation at temperature $T$. Calculating the system's entropy in any given macrostate, then, requires us to 
allow for the electromagnetic degrees of freedom as well as those of the charged particles.

In some situations --- the cores of stars, for instance --- the entropy of the black-body radiation is comparable in significance to the entropy of the matter. But in terrestrial plasmas at, say, a few tens of thousands of kelvin, the entropy of the matter in the plasma is overwhelmingly dominant.\footnote{To check this, recall (or note from section~\ref{uniformclouds}) that up to a constant (which does not affect thermodynamic behaviour), the entropy of an $N$-particle ideal gas in a volume $V$ is $Nk [\ln(V)+(3/2) \ln(T)]$, use this formula to get an order-of-magnitude estimate for the entropy of the matter in the plasma, and compare this to the entropy $(4/3)\sigma V T^3$ of black-body radiation (where $\sigma=(\pi^2/15)(k/\hbar c)^3 k\sim 10^{-16}JK^{-4}m^{-3}$ is the Stefan-Boltzmann constant). One can readily see that only at very low densities and/or high temperatures is the radiation term significant compared with the matter term.}

But this does not mean that electromagnetism is irrelevant to plasma thermodynamics in conditions less extreme than stellar cores. Even if the dynamical degrees of freedom of the electromagnetic field can be neglected, the presence of significant electromagnetic interactions will change the statistical mechanics of the plasma particles themselves. In an ideal gas, there are (\emph{ex hypothesi}) no interactions between the particles; in a plasma, there are strong short-range interactions. This does \emph{not} change the volume of any given macrostate (taking macrostates in the Boltzmannian sense as regions in phase space in which the macroscopic parameters of the system are constant or nearly so). It does, however, change the values of the macroscopic parameters characterising each macrostate (that is, each salient phase-space volume): in particular, the presence of short-range interactions will normally change the energy of various macrostates. And since the system is constrained to move only on a hypersurface of constant energy, this will have strong effects on the thermodynamics of the plasma, and in particular on which properties its equilibrium state has at a given energy.

In summary: introducing interactions to a system of particles affects the statistical mechanics of that system in two ways: via the additional degrees of freedom that may be associated with that interaction, and by the dynamical effects of the interaction on the macroscopic parameters associated to macrostates. And we should expect the same to be true of gravity. A gravitating system differs from a non-gravitating one both because of the degrees of freedom associated with the gravitational field (\iec, the spacetime metric) and because gravitational interactions change the dynamical properties of the system. 

However, there are very few physical situations in the Universe where there are any reasons to believe the entropy of the gravitational field to be relevant. Black holes, arguably, are counter-examples: the Schwarzschild solution is entirely matter free, and even in more realistic black holes, the matter that formed them is long out of sight behind the event horizon, leaving nothing but the geometry, yet there are good reasons to believe that black holes still have  (very high) entropy. (And this means that, were our concern the ``fourth principle'' of section~\ref{principles} --- the purported improbability of the intitial condition --- then gravitational entropy would be \emph{highly} relevant.)

 But in practice, in most physical situations  the dynamical degrees of freedom of gravity are frozen out, and we can treat gravity as simply a force between matter particles, without dynamical freedom of its own (indeed, for most purposes we can treat it as Newtonian). In particular, in classical cosmology there is no reason to think that significant gravitational radiation is present, and the motion of matter is far too slow to generate significant gravitational radiation. And the formation of black holes --- for all that it may have been a major contribution to the overall entropy increase of the Universe --- does not seem to have anything very much to do with the reasons why out-of-equilibrium processes can be found on Earth and its vicinity. It therefore seems reasonable to analyse the statistical mechanics of such situations --- and in particular, the statistical mechanics of cosmology --- without allowing for the entropic contribution of the gravitational field. 

(Incidentally, this seems rather to draw the sting from some recent --- and fairly scathing --- remarks about gravitational entropy by \citeN{earmanph}. Earman argues that talk of gravitational entropy is nonsense --- ``not even false'' --- because we have almost no understanding of how to calculate the statistical mechanics of general relativity, and may not have until we have obtained a quantum theory of gravity, and so any consideration of gravity in discussions of statistical is likewise nonsense. For this reason, he argues that sensible discussions of the Second Law cannot be held except in regions whose size is small enough that gravity may be neglected. But as we have just seen, there is more to the significance of gravity in statistical mechanics than is entailed simply by the phase-space volume associated with the microstate of the gravitational field.)

How, then, should we calculate the entropy of the early Universe? In a totally conventional way: work out the phase-space volume of any given macrostate, and take its logarithm. (Granted, doing this for an infinite universe may be awkward, but there does not seem to be any problem in principle in taking some fixed large region in the early Universe, calculating the entropy of that region, and then comparing it to the entropy of the matter in that region at earlier and later times. (Cosmologists routinely speak of the ``entropy per baryon'' as a natural measure of this quantity, since baryon number has been conserved in the Universe since at least times far earlier than those relevant here.) If gravity plays a relevant role here, it is only in that it changes the energy associated to any given macrostate, and so potentially affects the range of macrostates dynamically available, and in turn the statistical-mechanical behaviour.

The obvious question, then, is: \emph{how} does it in fact affect that behaviour? 
It is to this question that I now turn.

\section{Entropy of uniform self-gravitating gas clouds}\label{uniformclouds}

Heuristically, it is easy to see how gravity \emph{might} make the entropy of condensed systems higher, rather than lower, than that of more diffuse systems. The first step is to recognise that what is really relevant is whether a condensed system is higher in entropy than a diffuse system \emph{of the same total energy}. (Of course there are some diffuse systems higher in entropy than some condensed systems, and vice versa: we can achieve this by heating up the system that we want to be higher in entropy!) 

The second step is to notice that the entropy of a system of particles, being a measure of phase-space volume, depends on both the typical momentum of a particle (and thus ultimately on the sysem's total kinetic energy) and on the volume accessible to the particle (and thus ultimately on the system's total volume). Hotter systems have higher entropy; smaller systems (given fixed particle number) have lower entropy.

In non-gravitating systems (and here I mean really to consider dilute gases), typically the inter-particle forces are either negligible, or short-range and repulsive. This means that free expansion is a win-win situation for such systems, as far as entropy is concerned: not only does the volume increase, but the potential energy of the system is either negligible or actually decreases as the repulsive forces do work on the particles, and so conservation of energy means that the kinetic energy either increases or at least remains constant. No 
wonder, then, that in the absence of gravity, expansion is an entropy-increasing process; no wonder that in the absence of gravity, gases expand to fill the space available to them; no wonder that the equation of state of a gas at equilibrium depends on its volume as well as its energy. 

Gravity changes things. Expansion of a self-gravitating gas (definitionally) increases the gas's  volume (which increases the entropy), but it also increases the potential energy and thus decreases the kinetic energy, as particles must do work against the attractive gravitational field. So we should expect expanding gases to cool down, and this decreases the entropy. (Or conversely, and in more microphysical terms, if a gravitating gas contracts then the spatial volume available to particles decreases but the momentum-space volume increases.) In principle, it could be the case that the cooling effect is entropically more important than the expansion, so that contraction is entropy-increasing. (Indeed, this seems to be precisely what Albert had in mind in the quotation in section~\ref{principles}.)

Of course, such qualitative ideas need to be examined quantitatively, and we can do so by means of a simple model. Consider an ideal, uniform gas of total energy $E$ and kinetic energy $K$, containing $N$ particles whose internal degrees of freedom I will assume to be either nonexistent or irrelevant at the system's current temperatures, and contained within a spherical region of radius $R$. Equilibrium thermodynamics tells us\footnote{See \citeN[p.127]{garrod}, or any other textbook on statistical mechanics.} that the entropy of this system is 
\begin{equation}
S(E,R)=N k_B\left(  \ln (V/N)+\frac{3}{2}\ln (K/N)+C\right)\label{gasentropy1}
\end{equation}
\begin{equation}
=N k_B\left( 3 \ln (R/N)+\frac{3}{2}\ln (K/N)+C'\right),
\end{equation}
where $k_B$ is Boltzmann's constant, $V=4\pi R^3/3$ is the gas volume, and $C$, $C'$ are constants which will be irrelevant for our purposes. 

To understand this equation as a function of $E$ and $R$ (rather than just $K$ and $R$) we need to know the relationship between total and kinetic energy. In the absence of gravity (or any other significant inter-particle interactions) this is straightforward: $E\simeq K$. 

If the system has signicant self-gravity, though, we also need to consider the potential energy $U$, which Newton's theory tells us is given by
\begin{equation}
U \sim - \frac{G (N \mu)^2}{R}
\end{equation}
where $\mu$ is the mass per particle
(the general form of this equation follows from dimensional analysis; the proportionality constant is actually $3/5$, as can be checked by calculus, but it would be committing the sin of false precision for me to suggest that this level of accuracy can be relevant to my simple example!). It will actually be useful later, though, to consider the more general potential
\begin{equation}
U \sim - \frac{G (N \mu)^2}{R^\alpha}.
\end{equation}
And of course conservation of energy tells us that
\begin{equation}
K=E-U
\end{equation}
so that we can once again regard the entropy of the gas cloud as a function $S(E,R)$ of $E$ and $R$.

How does the entropy change when the radius changes? Elementary calculus tells us that
\begin{equation}
\dbd{S}{R}= 3 N k_B \left( \frac{1}{R}+\frac{1}{2K}\dbd{K}{R}\right)
\end{equation}
and, since $K=E-U$, this can be rewritten as
\begin{equation}
\dbd{S}{R}= 3 N k_B \left( \frac{1}{R}-\frac{1}{2K}\dbd{U}{R}\right)
=
\frac{3 N k_B}{R} \left( 1+\frac{\alpha U}{2K}\right).
\end{equation}
The entropy, therefore, will have an extremum when $\alpha U=-2K$ (and we can readily establish that it is in fact a maximum). This allows us to solve for $U$,  $K$, and $R$ in terms of $E$; we find that 
\begin{equation}
U=\frac{2 E}{2-\alpha};\,\,\,\, K=-\frac{\alpha E}{2-\alpha};\,\,\,\, 
R^\alpha=\frac{\alpha-2}{2}\frac{G(N\mu)^2}{E}.
\end{equation}
In particular, when $\alpha=1$ (the Newtonian case), $U=2E$, $K=-E$, and $R\sim -1/E$. We conclude that the maximum-entropy state for a uniform self-gravitating gas under Newtonian gravity and with negative total energy $E$ is one where the magnitude of the potential energy is twice the kinetic energy.\footnote{This is actually a special case of a much more general result in classical mechanics: the virial theorem. See \citeN[pp.\,358--364]{binneytremaine} for a discussion.}

So: suppose we consider a large, cold gas cloud. If the total energy of the cloud is positive, expansion will always increase its entropy; this is probably unsurprising, as the average velocity of the gas particles exceeds the escape velocity. If it is negative, it may be entropically favourable for the cloud to contract somewhat, but the effect is not normally that marked: for instance, on this model even a cloud which begins at absolute zero will only contract to half its initial radius before reaching its maximum-entropy state.

Note that, were the gravitational force stronger, contraction would be more strongly favoured: in a universe with an inverse-cube gravitational force (so that $\alpha=2$, there would be no maximum-entropy state and it would be entropicly favourable for uniform gas clouds to contract indefinitely. In particular, this gives us (admittedly only heuristic) grounds to expect that the story is different in strongly relativistic regimes, where the effects of gravity are significantly stronger than in the Newtonian limit (such as the collapse of neutron stars to form black holes).

However, and as has already been stressed, in most physically relevant situations in astrophysics and cosmology the Newtonian approximation works fine. This being the case, our model --- for all its simplicity --- seems to give us some grounds for scepticism as to whether gravitational collapse is really an entropy-increasing process. And we can gain direct evidence for this by using the above formula for the entropy of ideal gases to get a very crude estimate of the entropy of the matter in the Universe now and at earlier epochs. For instance, about 300,000 years after the Big Bang (the last point at which the microwave background radiation interacted significantly with baryonic matter), the baryon density was $\sim 10^{14}$ baryons $/\mathrm{m}^3$ and the temperature was $\sim 10^4 \mathrm{K} $.  Presently, most of the baryonic matter is locked up in stars of typical temperature $\sim 10^7 \mathrm{K}$ and typical density $\sim 10^{30}$ 
baryons $/\mathrm{m}^3$ (and the number of matter particles has not changed dramatically in the interim). It follows from equation \ref{gasentropy1} that the difference in entropy per baryon between ideal gases of densities $\rho_1,\rho_2$ and temperatures $T_1,T_2$ is 
\begin{equation}
\Delta S =k_B\left( - \ln (\rho_2/\rho_1)+\frac{3}{2}\ln (T_2/T_1)\right)\sim -25 k_B.
\end{equation}
That is: the entropy of the baryonic matter in the Universe is much lower now than it was in much earlier epochs. (The calculation is obviously extremely crude, but the effect should be large enough to assuage concerns about this.)
This empirical data-point only serves to underpin the theoretical concern with which I began the paper (the Initial State Problem): how can the entropy of the early Universe be higher than the entropy of the present Universe? To gain a satisfactory understanding of this, we need to return to the gravitating gas cloud model and reconsider an important assumption of that model: that the maximum-entropy state of the cloud is uniform.

\section{The gravothermal catastrophe}

It is no mystery why uniformity is the maximum-entropy state of a non-gravitating gas. Suppose, for instance, that we consider an initially uniform gas in a box, and imagine some small heat flow from the left hand side to the right. This would cause the left hand side to cool and the right hand side to heat up, which is a process that overall reduces entropy. Similarly, if we imagine that the left hand side contracts a bit and the right hand side expands a bit, this too will lead to a decrease in overall entropy, with the reduction in phase space volume of the left hand side being (slightly) higner than the increase in phase space volume of the right hand side.

But in a gravitationally bound gas, the energy and volume of the gas are not independent parameters: the gas is not contained in a box of fixed size, but is able to expand and contract freely provided energy is conserved in the process. We have already seen, for instance, that on the assumption that the gas remains uniform --- an assumption we are about to lift --- such a gas has an equilibrium radius proportional to the reciprocal of (the magnitude of) its energy.

Now, however, suppose we consider a spherical gas cloud broken into two concentric parts, a `core' and `envelope', each of which we assume to be uniform. We have already seen that uniform gravitating gases maximise entropy when they satisfy $U=2E, K=-E$ (where $K,U,E$ are respectively the kinetic, potential and total energies), so we will make that assumption for each of the core and the envelope.

Now suppose that there is some small heat flow from the core to the envelope. In the immediate term this reduces the kinetic energy of the core, but it also reduces the total energy of the core by the same amount, so in fact it will be entropically favourable for the core to contract ($U=2E$ and $E$ has just become more negative) and heat up ($K=-E$). Somewhat counter-intuitively, gravitating systems increase in temperature when they emit heat (for this reason they are often said to have negative heat capacities; see Callender~\citeyear{callenderpast,callenderheavy} for further discussion of this and other anomalous features of gravitational statistical mechanics). We have already seen that this process is entropy-decreasing: even if the contraction occurs without heat transfer, the drop in entropy from the decrease in volume more than offsets the increase in entropy from the rising temperature, and in this case (where there is heat loss) the effect will be even more pronounced.

However, to appreciate the total entropy budget of the cloud we need to allow for the effect of the heat transfer to the envelope. Conservation of energy means that the envelope now has higher (that is, less negative) energy than before; as such, it is entropically favourable for it to cool down and expand, and this process is entropy-\emph{in}creasing.

If core and envelope begin at the same temperature, then to first order the overall process must conserve entropy. (Temperature is $\delta Q/\delta S$, by definition.) But once a small amount of energy has been transferred from core to envelope, the core temperature will have gone up and the envelope temperature will have gone down. The process, therefore, will be self-perpetuating: initial temperature gradients will be increased, not dissipated, by heat flow.

It follows that an initially uniform cloud of gravitating particles will not remain that way. The cloud will evolve into states where the core gets more and more concentrated and the envelope gets more and more diffuse, and the process, in principle, will continue without limit.

This result is known in astrophysics as the \emph{gravothermal catastrophe},\footnote{See \citeN[pp.\,567--572]{binneytremaine} for a discussion, and for further references.} and it can be established in much more generality than for my simple model. It is, in fact, realised in nature by globular clusters (the very dense, spherical clusters of $\sim 10^6$ old stars found in the galactic halo, which appear to undergo precisely this evolution --- at least until they reach a stage where very close encounters between individual stars allows the formation of closely bound binary systems, freeing up energy to keep the cluster stable for a while longer.\footnote{For general information about globular clusters, see~\citeN[pp.\,327--376]{binneymerrifield}; for a discussion of their dynamics, see~\citeN[pp.\,596--632]{binneytremaine}.}

For systems of gas and dust, things are slightly different --- the same logic as before tells us that is entropically favourable for them to contract by emitting heat to the environment, but the most effective channel for \emph{them} to do so is via electromagnetic radiation. In this way, the entropy of the radiation in the Universe increases, more than compensating for the decrease in entropy of the contracting cloud.\footnote{Why don't globular clusters radiate? Because radiative emissions occur when matter interacts electromagnetically and thus accelerates. Stars are not charged (or are negligibly charged) and electromagnetic interactions between them are negligible. In stellar systems with considerable amounts of interstellar gas and dust, however, stars can radiate indirectly when their passage through this interstellar medium transfers energy to it. But globular clusters don't contain much gas.}

A corollary of this is that gravity-dominated systems do not have equilibrium states, unless we count black holes. A system dominated by gravitational interactions --- such as a cloud of interstellar gas, or a star cluster --- will contract (or at any rate its core will contract) and emit heat until either some new source of energy is found to replace the heat being transmitted to the environment, or some non-gravitational effects come into play to resist the attractive force of gravity.\footnote{I say ``effects'' rather than ``forces'' advisedly: in practice the most important astrophysical effect of this type is the pressure caused by the Pauli exclusion principle. This phenomenon is responsible for the stability of white dwarves and neutron stars.} In stars, for instance, it is nuclear fusion which provides the energy source; in globular clusters, it is the formation of hard binaries (strictly speaking the latter is another gravitational energy source). But eventually this energy source will be exhausted and gravitational collapse will resume. 

\section{Black hole entropy as a special case}

We can now see that black holes, often cited as the paradigm case of where gravitational effects make clumped systems high-entropy, are actually something of a special case. In general, highly clumped systems actually have very low-entropy, with the clumping process being entropically favourable only because of the energy generated by gravitational collapse and dispersed as heat. But black holes appear to have extraordinarily high entropy.

To illustrate this, consider another model.\footnote{Here and elsewhere I omit detailed references for standard data and results from stellar astrophysics; see, \egc, \citeN{salariscassisi} and references therein for a fuller discussion.} Take two or three solar masses of matter (the sun weighs $2 \times 10^{30}$ kilograms) and scatter it uniformly through a box that is (say) a hundred light years on a side. The matter will contract, heat up, and radiate: in doing so, its entropy will drop and the box will come to contain high-entropy black-body radiation. As the matter contracts to stellar densities, fusion reactions would normally occur; let us suppose, though, that the matter in the box is iron (which does not undergo exothermic nuclear reactions), so that the cloud just keeps contracting past those densities. In doing so, the particles in the cloud will become so compressed together that electrons and protons combine to form neutrons. Only when the star is comprised of neutron matter --- with a density of $\sim 10^{14} \mathrm{kgm}^{-3}$ --- will its heat capacity become positive and its collapse halt. At this point it will have a temperature of $\sim 10^{12}\mathrm{K}$, but it will now be entropically favourable for it to cool by emitting radiation\footnote{Actually most of the early cooling will occur by neutrino emission; ignore this, for the sake of the model!}.

At this point, the collapse of the matter has released perhaps a tenth of its mass of radiation,\footnote{This is a ball-park estimate based on the assumption that the star's binding energy can be calculated using Newtonian gravity; nothing hangs on its actual magnitude.} or $\sim 10^{46}$ joules. Black-body radiation at temperature $T$ has energy $E=\sigma V T^4$, where $V$ is the volume of the box and $\sigma=\mathrm{(2 \pi^5 k_B^4/15 c^3 h^3)}\sim 10^{-16} \mathrm{J m^{-3}K^{-4}}$, so the temperature of the radiation works out as $\sim 100$ K. The cooling of the neutron star will increase this somewhat, but the total energy emitted in the cooling will be of the same order as the energy emitted in collapse, so the system will eventually come into equilibrium with star and radiation alike at a temperature of $\sim 100 \mathrm{K}$.

At this point, the neutron star is a cold, highly degenerate ball of neutron matter. A rough estimate for its entropy per neutron is $\sim k_B T ( T/T_F)$, where $E_F \sim 10^{9} \mathrm{K}$ is the Fermi temperature of the star:\footnote{The Fermi temperature of an ideal gas of fermions is $(\hbar^2/m k_B)(\sim 3\pi^2 N)^{2/3}$, where $N$ and $m$ are the number density and the mass. The energy of a low-temperature Fermi gas is $E\sim E_0 +k_B(T/T_F)^2$, and $\Delta S \sim \Delta U/\Delta T$.} this tells us that the star has entropy about $10^{-5}k_B$ per neutron, or about $10^{29}\mathrm{JK^{-1}}$ in total. 

The entropy of black-body radiation is given by $S=4 E/3T$. So at this time, the entropy of the radiation in the box is $\sim 10^{45} \mathrm{JK^{-1}}$. In other words, the entropy of the radiation is sixteen orders of magnitude higher than that of the star. (The estimate is very crude, but sixteen orders of magnitude can cope with a lot of crudeness!)

Now suppose that we add just a little more matter to the star. There is a critical mass around 3 solar masses at which neutron matter is not stable against gravitational pressure.\footnote{Calculating the exact value of the critical mass is impossible at present since the structure of neutron matter is poorly understood.} If the box initially contained just less than this critical mass, adding just a small amount of extra matter will cause the star to collapse into a black hole.

This process will of course cause additional radiation, further heating the black-body radiation in the box. But this effect is dwarfed by the entropy change of the star. The entropy and temperature of a black hole of mass $M$ are\footnote{See \citeN{waldqft} for a review of the thermodynamics of black holes.}
\begin{equation}
S=\frac{k_B \pi \mathrm{G}M^2}{\hbar \mathrm{c}},\,\,\,\,T=\frac{\hbar \mathrm{c^3}}{2k_B \pi \mathrm{G}M},
\end{equation}
and so after the collapse the star's temperature decreases from $\sim 100$ K to $\sim 10^{-6}$ K, and its entropy increases by more than thirty orders of magnitude, from $\sim 10^{29}\mathrm{JK^{-1}}$ to $\sim 10^{62} \mathrm{JK^{-1}}$.

At this point, the radiation field is overwhelmingly hotter than the black hole, and radiation absorption by the hole will be much more rapid than emission of Hawking radiation by the hole. Furthermore, black holes have negative heat capacity, so this absorption only serves to cool the hole down. So when the system is in equilibrium again, almost all of the energy emitted in the original 
collapse of the matter in the box has been reabsorbed by the black hole, and all that is left is a very weak radiation field at a temperature of $\sim 10^{-6}$ K and thus an entropy of only $\sim 10^{21} \mathrm{JK^{-1}}$.

So it is quite misleading to see the process of black hole formation as simply the limiting case of gravitational clumping leading to higher entropy. In the normal case, the process is driven by heating of the environment by the collapsing system; a significant amount of energy is deposited in that environment and it has the overwhelming majority of the total entropy. When black holes form, it is the collapsing object itself which has the high entropy, and (in the very long run) the environment is actually cooled down by the collapse. 

\section{Entropy increase from gravitation and nuclear fusion}

We are now in a position to understand how the entropy of matter in stars in the present epoch can be lower (per baryon) than in the early universe, without a clash with the 2nd law. Namely, as matter contracted and clumped, the clumping process caused it to emit heat (despite also increasing in temperature) and that heat was emitted in the form of very-high-entropy radiation. So the entropy decrease in the entropy of matter is more than compensated for by the entropy increase in the entropy of the electromagnetic field.

Indeed, it might appear that a full solution of our Initial State Problem is at hand. The early universe was indeed at local thermal equilibrium, but due to the possibility of gravitational clumping this was not a global equilibrium: higher-entropy states were available at the same energy in which the matter was concentrated into the form of stars and the liberated gravitational potential energy was broadcast into space as high-entropy radiation. The universe has now moved into such a state; ergo, its entropy has increased.

But there is a problem with this story. The typical star takes $\sim 10^7$ years to contract gravitationally from a dispersed cloud (emitting radiation, and thus increasing global entropy, in the process), but then continues to shine for $\sim 10^{10}$ more years, powered by the fusion of hydrogen to helium, without appreciably changing in size. And the emission power (that is, radiation emitted per unit time) in the second phase is comparable in magnitude to the maximum emission power in the first phase.\footnote{The emission rate of a star is basically just a function of its size and mass, and is independent of whatever internal process, if any, is replenishing lost energy. So a protostar contracts, heats up, and radiates until its core gets hot enough for fusion power to generate as much energy as the star is losing from radiation; then it remains at that size until its fuel is exhausted.} So actually, over a star's lifetime most of the entropy increase due to the star is facilitated by nuclear reactions and not by gravitational collapse. And so it must follow that the material making up a star is not even at \emph{local} thermal equilibrium.

How can this be? The stellar core certainly looks like a good candidate for local thermal equilibrium: it is a very hot, very dense, plasma, made of a mixture of protons, electrons, and deuterium and helium nuclei. But in fact, the very existence of fusion reactions shows that it is actually entropically favourable for the protons to fuse to helium, and so the plasma must not be in thermal equilibrium after all.

An analogy may help: a gas comprising a mixture of hydrogen and oxygen, both at room temperature and atmospheric pressure, may appear to be at thermal equilibrium. In fact, though, it is \emph{highly} entropically favourable for the gases to react to form water. But the energy barrier to doing so is high, so that the gases will only obtain this higher-entropy state on extremely long timescales. On the other hand, intermolecular collisions will very rapidly redistribute kinetic energy amongst the gas molecules. So even on fairly short timescales, it makes sense to treat the spatial degrees of freedom of the gases as being at equilibrium (and, hence, to treat each gas as separately at equilibrium) but it does not make sense to assume that the mixture \emph{as a whole} is at thermal equilibrium. If the system is altered so as to make the energy barrier preventing the reaction a bit lower --- for example, by striking a match --- then the mixture will rapidly move to true thermal equilibrium.\footnote{Or tru\emph{er} thermal equilibrium, at any rate; see below.}

It will be helpful for us to understand in a bit more detail exactly why the fusion reaction in the sun is entropically favourable. In general, any exothermic (that is, energy-generating) fusion reaction (be it nuclear fusion, or the fusion of hydrogen and oxygen to create water) can decrease or increase entropy, depending on which of two processes is dominant. Firstly, there is an entropy decrease caused by fusion because the number of particles in the system has decreased and so the phase space volume available to them has decreased. Secondly, there is an entropy increase because the reaction is exothermic and the energy released will heat up the system (or be radiated as high-entropy radiation). Which of these processes will dominate will depend on how hot the system is: the hotter it is, the less difference will be made to the entropy by the energy release relative to the difference made by the particle number decrease. We therefore expect that at low temperatures, exothermic fusion reactions will be entropically favourable; at high temperatures, they will be entropically unfavourable. 

This is familiar in the case of the hydrogen-oxygen system we have just considered. At room temperature, the reaction is entropically favoured; but heat up water to a sufficiently high temperature, and it dissociates into hydrogen and oxygen. It may seem counter-intuitive in the case of stellar fusion, though: if fusion is entropically favoured at high but not at low temperatures, why does it occur in the stellar core but not on Earth's surface? The answer is two-fold. Firstly, the core of the sun may be hot by terrestrial standards, but not relative to the energies generated by fusion; secondly, although fusion reactions are entropically favoured for terrestrial matter, the energy barrier is so colossal that in practice, it never happens.

This is really just the same behaviour that we have seen for hydrogen and oxygen. The reaction between them, although strongly favoured entropically, in practice never happens without some catalysing process (a spark, typically). In fact, the true local equilibrium state of the hydrogen-oxygen mixture (at given volume and energy) is not water vapour at all: it is iron plasma, created when the oxygen and hydrogen nuclei react to form iron nuclei. But the energy barrier for the reaction is so phenomenal that one would have to wait essentially forever before the system reached that equilibrium. 

Although hydrogen plasma is not a local equilibrium state in the cold present-day universe --- even in relatively less frigid places like the cores of stars --- it was so in the early Universe. As a crude estimate, the exothermic nature of the hydrogen-helium fusion reaction can be expected to be negligible in the thermodynamics of plasma when the energy release is small compared to typical hydrogen-nucleus kinetic energies. Since the energy release in question is about one percent of the rest mass $m_p$ of the proton, this means that if
\begin{equation}
k_B T \gg 0.01 m_p c^2
\end{equation}
--- and therefore if $T \gg 10^{11} \mathrm{K}$ we expect hydrogen plasma to be a local thermal equilibrium state. Such temperatures were obtained a few minutes  after the Big Bang; at that time, the matter in the universe is believed to have been in the form of protons, neutrons and electrons, and to have been at local thermal equilibrium.\footnote{Here and afterwards I omit detailed references for cosmology; for details see, \egc, \citeN{weinbergcosmology} and references therein. \citeN{arnett} provides a good discussion of cosmological nucleosynthesis.}

So how did the Universe get out of local thermal equilibrium? As it expanded, for a while it remained in thermal equilibrium, but by the point at which its temperature approached $10^{11} \mathrm{K}$, the energy barriers to fusion became high enough relative to typical particle energies that the fusion reactions required to maintain equilibrium were unable to occur fast enough. So only about a quarter of the protons and neutrons present underwent fusion (and then, mostly speaking, only to helium). The bulk of the baryonic matter in the universe remained as free protons (and free neutrons, which quickly decayed into free protons), even though this state was no longer a thermal equilibrium state. Only the later concentration of interstellar gas into stars allowed fusion to resume.

\section{Thermodynamics in an expanding Universe}

This is all very well, the reader may be thinking, but a central problem seems to have been ignored. Recall the Initial State problem: how can the early Universe have been in local thermal equilibrium, given that it is out of equilibrium now? Taking gravity into account had the potential to have resolved this problem: because of the attractive nature of the gravitational force, systems can be in thermal equilibrium locally but not globally. But we have just seen that the universe came out of local thermal equilibrium at a time at which the matter distribution remained essentially uniform, so this resolution is not available to us, and the original problem remains.

In fact, the solution has nothing to do with gravitational clustering. There is another, and far simpler, reason why the early universe was not and cannot have been in thermal equilibrium. Equilibrium, by definition, is a state in which the macroscopic parameters of the system are time-independent \ldots

\ldots and the Universe is expanding.

To see just what the significance is of the expansion of the Universe, it will be helpful to consider another simple model. Suppose we confine a gas in a cylinder with a frictionless, heatproof piston at one end. Initially, the gas is at equilibrium at some energy $U$ and volume $V$, and these parameters serve jointly to determine the entropy $S(U,V)$. If the piston is pulled partially out, the volume will expand to $V'$, and the energy will decrease to $U'$ --- decrease, because the gas molecules do work on the piston as it is pulled out. The exact value of $U'$ cannot be calculated via equilibrium thermodynamics and will depend on the details of how the piston is moved. But we do know, from the second law, that $S(U',V')\geq S(U,V)$.

There is a special case in which equilibrium thermodynamics tells us rather more than that. If the piston is pulled out sufficiently slowly, we will be able to treat the gas as being effectively instantaneously in equilibrium. (``Sufficiently slowly'' must therefore mean ``the volume changes slowly compared to the timescales on which molecular collisions cause the gas to equilibrate''.)
In this idealisation, we say that the gas is expanding \emph{quasi-statically}, and it is a standard result of equilibrium thermodynamics that a system expanding quasistatically and transmitting no heat to the environment must have a constant entropy. 

So, if we quasi-statically pull out the piston and then quasi-statically push it back in to the same point, the volume of the system will increase to $V'$ and then decrease to $V$ again. Since the entropy remains constant over this period, it follows that the energy will decrease from $U$ to some energy $U'$ dependent only on $V,V'$ and $U$, and will then increase back to $U$: the work done on the piston by the gas in the expansion phase exactly cancels the work done on the gas by the piston in the contraction phase.

On the other hand, if the piston is pulled out and pushed back a little more quickly than this, the entropy will increase in both the expansion and contraction; hence, the system energy will be higher after this process than before, even though the volume is unchanged. The extra energy has come from the person working the piston: the pressure was higher, on average, in the contraction than the expansion phase, so that the work done on the piston by the gas is less than the work done on the gas by the piston.

Now suppose that the ``gas'' is actually plasma at the kinds of temperatures and pressures found in the early Universe (and that the box, the piston, and the muscles of the person working the piston are made of something truly remarkable!) The story is essentially unchanged by this substitution: if the box is expanded so much that the contents' density reduces to that of intergalactic gas, and then contracted back to its original volume, then provided the expansion and contraction are slow enough to be quasi-static, the energy and entropy will be unchanged; if they are faster than this, the contents of the box will have heated up and the energy increase equals the net work done on the box contents by the piston. 

The important point, though, is that the expansion and contraction must be slow enough that the box contents are in instantaneous equilibrium, and this means they must be slow not only relative to the timescales on which kinetic energy is redistributed by collisions between particles but also relative to the timescales on which fusion and fission reactions occur. At various points (notably as the box contents cool below $\sim 10^{11} \mathrm{K}$), this will require the expansion to be \emph{very} slow, much slower than the rate at which matter expanded and cooled in the actual early Universe. If the expansion is not this slow, the contents of the box will cease to be in instantaneous equilibrium, and the entropy of the box will rise. When the box is eventually compressed back to its original size, its contents will be hotter and higher-entropy than at the start of the process, and the piston will have done net work on the contents.

How does this relate to the real Universe? If we consider a cube of early-universe matter, and if we assume that the Universe is homogenous and isotropic (a good approximation until long after the periods of interest to us here), we can treat the expansion of this cube as adiabatic: homogeneity and isotropy rule out any net heat flow across its borders. In the early Universe, the expansion was also quasi-static, because the rate at which the overall size of the cube changed was very slow compared to the speed of equilibrating processes (\iec, collisions between particles) in the material of the cube. So to an extremely good approximation, in this phase of the Universe's expansion its entropy remained constant and its contents remained in instantaneous thermal equilibrium, cooling just enough to keep their entropy constant despite their volume increase. 

But once the Universe was cool enough that fusion reactions became entropically favourable, the relevant timescales changed. Now, for the expansion to preserve equilibrium, it had to be slow relative not only to interatomic collision timescales but to fusion timescales --- and these were considerably slower, so much so in fact that the quasi-static assumption just failed, and the matter in the cube came out of thermal equilibrium, remaining (for the most part) as protons, neutrons and electrons even though it would have been energetically favourable for it to undergo fusion.

What does this mean for entropy? Imagine, for simplicity, that the fusion processes just did not occur at all in the early Universe. Then the \emph{actual} entropy of the Universe would have remained constant during the 
disequilibration period (if fusion cannot occur at all, then the early-Universe plasma just behaves like a system whose equilibrium state does not involve fusion --- and the expansion is quasi-static for such systems). If some fusion reactions then do occur, those reactions will be straightforwardly entropy-increasing. In this sense, the rapidity of the expansion process does not increase the actual entropy, but it does increase the maximum possible entropy.

In fact, what actually happened in this phase of the Universe's expansion was that fusion reactions occurred partly but not wholly: about 25\% of the nucleons in the Universe were able to undergo fusion to helium before the densities and temperatures became too low. This process, then, was entropy-increasing, and thus irreversible.

To make this clear, suppose (\emph{contra} the current astrophysical consensus) that the Universe will contract again to a Big Crunch, and suppose (absurdly) that it had been \emph{so} perfectly uniform that no significant inhomogeneities had been able to develop over its lifetime. Then as the Universe contracted, and its contents warmed up, we would not see a reversal of the early-Universe fusion processes but rather a continuation of them: the fraction of nucleons bound into helium would not fall smoothly off from 25\% back to zero but would increase for a while, until the Universe became so hot that the nucleons broke up again.

It follows that the early-Universe fusion process was, straightforwardly, a Second Law process. And since it happened while the Universe was still virtually uniform, it acts as a plain counterexample to the thesis that the Second Law is only possible because of the effects of gravitational clustering.

It also follows that the collapsing Universe is hotter, for a given volume, than it was in its expansion phase --- just as was the simple box-and-piston model I discussed previously. In that model, the energy to heat the box came from the net work done in the expansion and contraction process. In cosmology, the increased energy of the matter in the Universe comes at the expense of the Universe's rate of expansion: the contraction phase is less rapid than the expansion phase. The expansion of the Universe is aided by the pressure of the matter in the Universe, and its contraction is hindered likewise. But since the contracting Universe is hotter than the expanding Universe, the latter effect is more pronounced than the former one. In a sense, then, the expansion process does `net work' on the matter in the Universe, although it is difficult to make this precise due to the problems involved in defining energy for the Universe as a whole.

In this light, it is helpful to consider an observation made by Penrose:
\begin{quote}
There is a common view that the entropy increase in the second law is somehow just a necessary consequence of the expansion of the universe \ldots This opinion seems to be based on the misunderstanding that there are comparatively few degrees of freedom available to the universe when it is `small', providing some kind of low `ceiling' to possible entropy values, and more available degrees of freedom when the universe gets larger, giving a higher `ceiling', thereby allowing higher entropies. As the universe expands, this allowable maximum would increase, so the actual entropy of the universe could increase also.
\ldots This cannot be the correct explanation for the entropy increase, for the degrees of freedom that are available to the universe are described by the total phase space $\mc{P}_U$. The dynamics of general relativity (which includes the degree of freedom defining the universe's size) are just as much described by the motion of [a] point in phase space $\mc{P}_U$ as are all the other physical processes involved. This phase space is just `there', and it does not in any sense `grow with time', time not being part of $\mc{P}_U$. \cite[p.701]{penroseroadtoreality}
\end{quote}
All this is dead right, of course. But equally, as soon as we recognise that the phase space of the Universe includes a degree of freedom corresponding to its overall size, we should also recognise that this means that the Universe is not at equilibrium after all. However, some subsystem of the Universe may be at equilibrium, and the interaction of that subsystem  with the other degrees of freedom may well increase the number of degrees of freedom available to it --- by increasing its energy, or its volume, or (as is in fact the case) by letting the latter increase more than enough to offset the effects of the former's decrease. The expansion of the universe does increase the available phase space volume for its small-scale degrees of freedom, not through the kinematic process of just getting bigger, but through the dynamical process of getting bigger too fast for the small-scale degrees of freedom to remain in instantaneous equilibrium.

\section{The real significance of gravity in thermodynamics}

If \emph{some} Second Law processes occurred in the early Universe whilst the matter distribution was close to being uniform, nonetheless \emph{most} of these processes would not have occurred without gravity. In particular, pretty much every entropy-increasing process on Earth relies ultimately on low-entropy material (be it sugar, oil, or uranium) produced by sunlight, and the Sun would not have formed without gravitational collapse.\footnote{The exceptions are processes which rely on fissionable or radioactive materials (either directly, as in nuclear power plants, or indirectly, as in geothermal energy, which is powered by radioisotopes in Earth's core) which were made in supernovae, and hydrogen bombs, which rely on the incompleteness of primordial fusion, so that plenty of hydrogen is still around.} So for all my discussion of thermonuclear entropy earlier, the reader may still have sympathy with the third Principle: that the Second Law --- perhaps excluding primordial fusion --- is ultimately bound up with gravity.

But I think this is misleading. Many things, including gravity are causally or explanatorily responsible for the Sun. But our interest here is not the whole problem of star formation, but specifically the thermodynamics of stars, and in particular the question of how the matter comprising stars starts off in a low-entropy state. Recall Penrose: 
\begin{quote}
It is the fact that this gas starts off as \emph{diffuse} that provides us with an enormous store of low entropy. We are still living off this store of low entropy, and will continue to do so for a long while to come. \cite{penroseenm}
\end{quote}
But the entropy-increasing processes of which we are living, at the moment, are not gravitational processes: they are nuclear fusion processes. And the low entropy that drives them is due not to the diffuse nature of the Sun but to the fact that is made of hydrogen and that this is not a local equilibrium state at the temperatures and pressures in the Sun's core.

Gravity, in fact, is important in to the Second Law as it plays out  on Earth not because gravitational entropy-increasing processes directly allow the creation of low-entropy matter on Earth, but because the non-gravitational entropy-increasing processes that do allow the creation of that matter have very high energy barriers which can only be crossed with the aid of gravity. In this sense, gravity is best thought of as a catalyst. Like the spark which allows a mixture of hydrogen and oxygen to ignite, gravity is essential to allowing the non-equilibrium matter in stars to move closer to local equilibrium. But the spark is of little relevance to the \emph{thermodynamics} of the ignition process. 

The point can be emphasised by means of another model --- one which is more or less realised in nature, in fact. Start with about two solar masses of hydrogen in a diffuse cloud and at a temperature of a few $\mathrm{K}$. If there were no gravity, this cloud would in practice remain in that state forever: it is entropically favourable for it to undergo fusion but the probability of any given collision leading to a fusion reaction is so fantastically low as to be irrelevant on any remotely significant timescale.\footnote{A crude estimate: if two nuclei collide with centre-of-mass energy $E$, the probability of a fusion reaction is $\sim \exp(-\sqrt{E_G/E})$, where $E_G$, the Gamow energy, is $493kEV$ for hydrogen fusion. A typical atom in a 100-Kelvin hydrogen gas has kinetic energy $\sim 0.01 eV$, and even the hottest atoms in a cloud of this size won't have energy more than a few dozen times this (from Maxwell's distribution law we expect the hottest atoms in an $N$-atom cloud to have energy $\sim \sqrt{N}$ times the typical energy. But even a $0.5ev$ collision has a probability $\exp(-10^3)\sim 10^{-10^3}$ of fusion reactions occurring. With a probability this small, the actual number of collisions per second doesn't really matter, and nor does the unit of time: the cloud will take at least $\sim 10^{1000}$ years to fuse to helium. } 

With gravity present, however, the gas will collapse and heat up, and eventually form a star. We will keep back $0.1$ solar masses for nefarious purposes to be revealed shortly; the remainder will contract until fusion occurs in its core and it becomes a star --- a yellow dwarf rather larger than the sun, in fact. After about $10^{10}$ years the star will exhaust its core fusion fuel and, after a brief period as a red giant star, it collapses to a ball of degenerate matter called a white dwarf, which consists of a mixture of carbon and oxygen nuclei supported against gravity by degeneracy pressure from its electrons. The white dwarf will weigh less than the original star because significant amounts of mass get blown into space in its solar wind during the late stages of its evolution; let us suppose that the remnant has a mass of about 1.4 solar masses.

1.4 solar masses is actually a fairly significant threshold: it is the so called ``Chandrasekhar limit'', the maximum mass that a ball of degenerate matter can attain before resuming gravitational collapse. So now let us pour on the 0.1 solar masses which we held back before. This will induce the star to begin collapsing, which will in turn heat it up until the carbon and oxygen nuclei in the star begin to fuse. For reasons that are not especially salient here, fusion under degenerate conditions is highly explosive, and the result is that the white dwarf explodes (or `disrupts', as astrophysicists euphemistically put it) in a gigantic explosion known as a thermonuclear supernova.\footnote{Supernovae come in two types: gravity-powered explosions of massive stars which leave a neutron star or black hole remnant, and thermonuclear explosions in smaller stars which leave no remnant. Thermonuclear supernovae are commonly called `Type Ia' supernovae, and core-collapse supernovae are called `Type II' supernovae --- but the type I/type II distinction is actually an observationally defined classification referring to the presence (type II) or absence (type I) of hydrogen lines in the supernova, and only roughly tracks the two theoretical kinds. See \citeN{arnett} for a detailed account of supernovae, and \citeN{salariscassisi} for an overview.}

Now consider: before the whole process began, the matter in the system was dispersed in a large, low-density cloud. And after the supernova --- which completely destroys the star, and leaves no residue whatever --- the matter is again dispersed in a large, low-density cloud. But the entropy of the system has increased: many of the protons in the original cloud now reside in larger nuclei, and this process has released a great deal of high-entropy radiation. The role of gravity in this process, pretty clearly, was to make it possible to happen in the first place, but there has been no net generation of entropy through gravitational collapse.

Now, none of this is to deny that gravitational processes often do generate significant entropy increases. Overwhelmingly the dominant entropy-increasing processes in the Universe are the formation of black holes; even setting this aside, significant amounts of the radiation in the Universe was generated by core-collapse supernovae, quasars, and the like, all of which are powered by gravity. But none of these processes have anything much to do with the playing out of the Second Law here on Earth.

It is perhaps helpful to imagine how the Universe might have been if the story in the third conventional principle, and in the associated conventional solution to the Initial State Problem, had been true. Suppose, for instance, that the initial state was a very uniform, very low-density cloud of iron atoms, and that the overall expansion or contraction of the universe was suppressed (say, by a carefully tuned cosmological constant). That universe truly would have been at local thermal equilibrium in its initial state. Small fluctuations in uniformity would then be magnified, over time, by the process of gravitational collapse: this (plausibly, at any rate) would eventually lead to the formation of stars, which would shine only through gravitational collapse and would eventually collapse to shrunken remnants. In the process, very significant energy flows would be created; these energy flows would often be able to kick systems out of local thermal equilibrium, allowing Second Law processes to occur on the surfaces of the planets orbiting these stars.

If our Universe had been like this, the third conventional principle, and the Conventional Solution, would have been pretty much exactly the right story to tell about the second law of thermodynamics. But our Universe is actually very different from this.

\section{Conclusion}

I began this paper with three Conventional Principles and one Conventional Solution, which I will restate here.

\begin{description}
\item[First conventional principle:] ``Taking account of gravity'' in a discussion of entropy means allowing for ``gravitational entropy'', the entropy of the gravitational degrees of freedom. In particular, the early Universe had very high entropy in its non-gravitational degrees of freedom but very low gravitational entropy.
\item[Second conventional principle:] When we do take account of gravity, dispersed low-density systems have low entropy, and concentrated high-density systems have high entropy. The stupendous entropy of black holes is simply a limiting case of this process.
\item[Third conventional principle:] Without the entropy-increasing effect of gravitational clustering, there would be no 2nd Law of thermodynamics. In particular, systems in our local neighborhood which are out of equilibrium have got that way because the process of gravitational clustering has allowed their entropy to drop.
\item[Conventional solution to the Initial State problem:] Although the early Universe was at local thermal equilibrium, it was not at global thermal equilibrium because it was highly uniform and the process of becoming non-uniform is entropy-increasing when gravity is taken into account.
\end{description}

I hope it is now clear to the reader what is and is not correct in these pieces of conventional wisdom. I will end by stating, in parallel to the Conventional Principles, a set of revised Principles which, I hope, really do capture what is actually going on in the thermodynamics of the Universe.
\begin{description}
\item[First revised principle:] ``Taking account of gravity'' in a discussion of entropy means allowing both for ``gravitational entropy'', the entropy of the gravitational degrees of freedom, and for the dynamical effects of gravity on the energy levels of the material degrees of freedom. With the exception of black holes, the former is not relevant in known physical situations. In particular, highly uniform systems have very low entropy compared to the entropy they would have had if they had been rearranged into a much more uneven state whilst keeping their total energy constant, even when gravitational degrees of freedom are entropically irrelevant.
\item[Second revised principle:] When we do take account of gravity, dispersed low-density systems actually have higher entropy than they would if they were allowed to collapse under gravity. But the heat capacity ofstrongly self-gravitating systems is negative, and this allows them to offset this entropy decrease by transferring heat to their surroundings and in doing so increasing in temperature. The stupendous entropy of black holes is rather a special case: in this case, but not in other situations involving gravitational collapse, the entropy of the collapsing object is higher before collapse than after.
\item[Third revised principle:] The second law of thermodynamics does not rely on the entropy-increasing effect of gravitational clustering: in particular, primordial fusion was entropy-increasing but occurred before any significant gravitational clustering. Essentially all second-law processes in the present epoch do owe their existence to gravity, but for many of them --- including most that happen on Earth --- the actual entropy increase is due to thermonuclear processes, and these processes occur because stellar matter is not in local thermal equilibrium, and it got that way due to the rapidity of the expansion of the Universe. In these cases, the role of gravity is best understood as the catalyst that allows already-thermodynamically-favourable nuclear reactions to occur in a reasonable length of time.
\item[Revised solution to the Initial State problem:] Although the early Universe was at local thermal equilibrium, it was not at global thermal equilibrium (a) because it was highly uniform and the process of becoming non-uniform is entropy-increasing when gravity is taken into account, and (b) because the Universe is expanding, and hence cannot have been at thermal equilibrium even when the possibility of non-uniformity is ignored. In both (a) and (b), a subsystem of the Universe (its small-scale degrees of freedom) is in thermal equilibrium, but the interactions between those degrees of freedom and other non-equilibrium degrees of freedom --- whether those describing the departure from uniformity or those describing overall expansion --- can cause the small-scale degrees of freedom to get out of equilibrium and thus to increase in entropy.
\end{description}

\section*{Acknowledgements}

I would like to thank Frank Arntzenius, Harvey Brown, Craig Callender, Roger Penrose, Simon Saunders, and Chris Timpson for useful comments.

\end{document}